\documentclass{article}
\title{\bf 4-D homogeneous isotropic cosmological models generated by
the 5-D vacuum.}
\author{\it  Vladimirov Yu.S.\thanks{e-mail: vlad@fund.phys.msu.su},
Kokarev S.S.\thanks{e-mail: sergey@yspu.yar.ru}
\\ Moscow State University,
\\ Yaroslavl State Pedagogical University}
\date{}
\begin{document}
\maketitle
\begin{abstract}

4-dimensional homogeneous isotropic cosmological models
obtained from solutions of vacuum 5-dimensional Einstein equations
are considered.
It is assumed, that the $G_{55}$--component of the 5-d metric simulates
matter in the comoving frame of reference. Observable 4-d metric is
defined up to conformal transformations of the
metric of 4-d section $\tilde{g}_{\mu\nu}$,
with a conformal factor  as a function of the component $G_{55}$.
It is demonstrated, that the form of this function
determines the matter equation of state. Possible equations of state are
analyzed separately for flat, open and close models.
\end{abstract}

\section{Introduction}

There is a number of fundamental problems in multidimensional geometric
Kaluza-Klein theories. Among such problems the question of physical meaning
of the scalar fields, given by extra components of metric $G_{55}, G_{66}$\dots,
takes a significant place. One can indicate four viewpoints on the status
of scalar fields:

1) Scalar fields are superfluous \cite{r1,p,pig,r13}. In this approach scalar
fields are eliminated on the geometric Lagrangian level by additional
conditions like $G_{55}=const$.

2) Extra components describe new unknown physical (massless) scalar fields
\cite{r3,brow,r4}. The problem of their experimental observation is posed.
Some researchers connect such fields with the hypothesis of the so-called
"fifth" force,
others with the hypothesis of variable fundamental constants. To this line
adjoin investigations in the context of scalar-tensor gravitation theories
of Jordan-Brans-Dicke type \cite{r5}.

3) Scalar fields can be understood as a geometric simulation of matter
\cite{wes1,wes2,wes3,wes4,wes5,wes6,wes7,wes8,r7,r10,r17}, which is substituted
arbitrarily to the
right-hand side of the Einstein equation through energy-momentum tensor $T_{\mu\nu}$
in 4-d GR.

4) In a number of variants of a unified multidimensional theory of the gravitational
and electroweak interactions geometric scalar fields describe Higgs particles
\cite{r8}.

In this paper the third from the above approaches is investigated
in the context of the 5-d Kaluza-Klein theory. The possibility of  describing
matter by a geometric scalar field in homogeneous isotropic cosmological
models of Friedmann type is considered.

\section{Physically meaningful quantities
in the 5-D Kaluza-Klein theory}

The problem of the physical meaning of scalar fields is closely related to
another question of greater importance: what quantities in
multidimensional theory are to be endowed by physical meaning?
Recall that the same question appears in 4-d GR. There it has been solved
with the help of the monad method of describing reference frames
\cite{r9,r10}, which is often called the 1+3-splitting method.
Therewith the metric can be represented as
\begin{equation}
g_{\mu\nu}=\tau_{\mu}\tau_{\nu}-h_{\mu\nu},
\end{equation}
where Greek indices take values 0,1,2,3; $\tau_{\mu}$ is 4-d
vector field, oriented along the corresponding observer's world line,
$h_{\mu\nu}$ is orthogonal to $\tau_{\mu}$ and has the meaning of a metric
in a local 3-d space section. A physical meaning is ascribed only to quantities
projected onto the $\tau$-direction or onto the orthogonal space section.

It is convenient to apply the monad method in a special coordinate system,
where the  coordinate lines $x^{i}=const,\ i=1,2,3$ are directed along world lines
of the frame of reference. In this case the vector $\tau_{\mu}$ has the components
\begin{equation}
\tau^{\mu}=\frac{g^{\mu}_{0}}{\sqrt{g_{00}}}.
\end{equation}
The monad method in such coordinate systems has been called method of chronometric
invariants \cite{r9}. In this method the following coordinate transformations
are dis\-tin\-gui\-shed:
\begin{equation}\label{chrt}
x'^{0}=x'^{0}(x^{0},x^{1},x^{2},x^{3});
\end{equation}
\begin{equation}\label{chrr}
x'^{i}=x'^{i}(x^{1},x^{2},x^{3}).
\end{equation}
A physical meaning is possessed by chronometrically invariant quantities, i.e. quantities,
which are invariant under  the transformations (\ref{chrt}) and covariant under the
transformations (\ref{chrr}).

For the description of physically meaningful quantities in 5-d Kaluza-Klein theory
one should  use a  similar method of 1+4-splitting of the 5-dimensional manifold
into the fifth direction and the 4-d classical space-time. The 5-d metric tensor
$G_{AB}$ has the following form:
\begin{equation}
G_{AB}=-\lambda_{A}\lambda_{B}+\tilde{g}_{AB},
\end{equation}
where $A,B=0,1,2,3,5,\ \lambda_{A}$ is a  5-d unit vector oriented along
the fifths direction, $\tilde g_{AB}$ is a metric of the 4-d section orthogonal to
$\lambda_{A}$. It appears reasonable that only quantities projected onto  the
$\lambda$-direction or orthogonal 4-d directions (by the projector
$\tilde g^{A}_{B}$) have a physical meaning.

There is a correspondence between the method of chronometric invariants and
that of gauge invariants\footnote{The appearance of this method can be
traced back to the works of Mandel \cite{r12}, then it was improved in the works
of Einstein and Bergmann \cite{r13}. It was ultimately formulated
in  the works by one of the present authors  \cite{r10}}
\cite{r10,r11} in 5-d theory, wherein coordinate systems,
in which $\lambda_{A}$ is oriented along the fifths coordinate
$x^{5}$ are selected. Then the vector $\lambda^{A}$ is given by the expression
\begin{equation}
\lambda^{A}=G^{A}_{5}/\sqrt{-G_{55}},
\end{equation}
the 4-d metric has the components
\begin{equation}
\tilde{g}_{\mu\nu}=G_{\mu\nu}-\frac{G_{5\mu}G_{5\nu}}{G_{55}};\ \
\tilde{g}^{\mu\nu}=G^{\mu\nu},
\end{equation}
and the following coordinate transformations are distinguished:
\begin{equation}\label{tr5}
x'^{5}=x'^{5}(x^{0},x^{1},x^{2},x^{3},x^{5});
\end{equation}
\begin{equation}\label{tr4}
x'^{\mu}=x'^{\mu}(x^{0},x^{1},x^{2},x^{3}).
\end{equation}

Gauge invariant quantities now become
physically meaningful, i.e.,
quantities invariant under transformations (\ref{tr5}) and
4-d covariant under the transformations (\ref{tr4}). Recall that in such a theory
the electromagnetic field tensor $F_{\mu\nu}$ is described by a gauge invariant
expression, while the gauge-non-invariant components  $\lambda_{\mu}$
correspond to  components of the electromagnetic vector potential.

A new circumstance, which was absent in 4-d GR, appears when describing the
"generalized frames of reference" of this kind. It is connected with the fact that
if the quantities of a multidimensional theory depend on extra coordinates, then
they will have an electric charge and may be other charges (hyphercharge, isospin).
Since the gravitational and electromagnetic fields are neutral, the 5-metric
should not dependent on $x^{5}$. Then (\ref{tr5}) should be
narrowed to
\begin{equation}\label{gtr5}
x'^{5}=x^{5}+f(x^{0},x^{1},x^{2},x^{3}),
\end{equation}
which conserves the cylindricity condition of $G_{AB}$ with respect to  $x^{5}$.
Under this transformations the components of $\lambda_{\mu}$ transform
by the law
\begin{equation}
\lambda'_{\mu}=\lambda_{\mu}+\frac{\partial f}{\partial x^{\mu}},
\end{equation}
that corresponds to the well-known gauge transformations of electrodynamics.

A new circumstance lies in the fact that the component $G_{55}$ is invariant
under the transformation (\ref{gtr5}). It means in turn that any of the
conformally corresponding metrics
\begin{equation}
g_{\mu\nu}=\tilde{g}_{\mu\nu}/{\cal F}(G_{55}),
\end{equation}
can be physically meaningful, where the conformal factor ${\cal F}(G_{55})$ is an
arbitrary function of $G_{55}$. There is an uncertainty in choosing
the conformal
factor. What factor ${\cal F}$ should be chosen? --- this question has
been posed and discussed in the book by one of the authors \cite{r10}.
A number of works on 5-d theories were there analyzed and the following
three most frequently used cases were distinguished:

a) ${\cal F}(G_{55})=1$ (immediate identification of the metric
obtained from the 1+4-splitting procedure). This case is used in most
works on 5-d theories, in particular, in the papers by Wesson \cite{wes1,wes4,wes5}.

b) The case ${\cal F}(G_{55})=(-G_{55})^{-1/2}\equiv1/\varphi$ is of interest
because after 1+4-splitting the effective energy-momentum tensor of the scalar field
has a canonical form (without second derivatives of $\varphi=\sqrt{-G_{55}})$.

c) The relation ${\cal F}(G_{55})=-G_{55}=\varphi^{2}$ has been used, for example,
in our works \cite{r11,r14}. This case is of interest because
the gravitational "constant" in the 4-d Einstein equations is a true
constant and in the
fifteenth Einstein equation the scalar of curvature ${}^{4}\!R$ appears with
the coefficient -1/6, as in the case of a conformally invariant scalar field.

Other cases are possible as well. A reasonable question arises:
what is hidden behind this arbitrary  choice of ${\cal F}(\varphi)$?
It turns out that, in the context of the approach to the physical meaning of $\varphi$
proposed here, a specific choice of the conformal factor means specifying
the equation matter of state.

\section {Exact solutions}

Let us take vacuum (without a right-hand side) 5-d Einstein equation
\begin{equation}\label{ein5}
{}^{5}R_{AB}=0.
\end{equation}
We shall find its solutions which correspond to 4-d homogeneous isotropic
cosmological models
\begin{equation}\label{fried5}
dI^{2}=dt^{2}-e^{2\lambda}[dr^{2}+\Sigma(r)(d\theta^{2}+\sin^{2}\theta
d\varphi^{2})]-e^{2\phi(t)}(dx^{5})^{2},
\end{equation}
where $\lambda(t)$ and $\phi(t)$ are two so far  unknown functions of $t$
and
\[
\Sigma(r)=\left\{
\begin{array}{ll}\label{curve}
r^{2}  & \mbox{for  flat 3-d section;} \\
\sin^{2}r & \mbox{for close 3-d section;} \\
\sinh^{2}r & \mbox{for open 3-d section.}
\end{array}
\right.
\]
Substituting (\ref{fried5}) to (\ref{ein5}), one can write the
Einstein equations in the form
\begin{equation}\label{ein5c}
\left\{
\begin{array}{l}
3(\ddot{\lambda}+\dot{\lambda}^{2})+\ddot{\phi}+\dot{\phi}^{2}=0;\\
\ddot{\phi}+\dot{\phi}^{2}+3\dot{\phi}\dot{\lambda}=0;\\
\ddot{\lambda}+3\dot{\lambda}^{2}+\dot{\lambda}\dot{\phi}+2se^{-2\lambda}=0,
\end{array}
\right.
\end{equation}
where $s=0,+1,-1$, according to the three outlined cases.

The solutions of (\ref{ein5c}) are exhausted by the following four:

1) The metric of Kasner type with a flat 3-d space section
\begin{equation}\label{flat1}
dI^{2}=dt^{2}-dr^{2}-r^{2}(d\theta^{2}+\sin^{2}\theta d\varphi^{2})-
t^{2}(dx^{5})^{2},
\end{equation}
has been considered in \cite{r17}. This solution describes flat 5-d space-time.

2) The metric with a flat 3-d section
\begin{equation}\label{flat2}
dI^{2}=dt^{2}-t[dr^{2}+r^{2}(d\theta^{2}+\sin^{2}\theta d\varphi^{2})]-
\frac{1}{t}d(x^{5})^{2},
\ \ (t\geq0)
\end{equation}
has been analyzed in \cite{wes1} and by others. It is of Kasner type too.

3) The metric with a 3-d space section of positive curvature
\begin{equation}\label{close}
dI^{2}=dt^{2}-(t^{2}_{0}-t^{2})[dr^{2}+\sin^{2}r(d\theta^{2}+
\sin^{2}\theta d\varphi^{2})]-\frac{t^{2}}{t^{2}_{0}-t^{2}}d(x^{5})^{2},
\end{equation}
\[(0\leq|t|\leq t_{0})\]
has been obtained in other coordinates in our work \cite{r10} and
by others \cite{iv}.

4) The metric with a 3-d space section of negative curvature
\begin{equation}\label{open}
dI^{2}=dt^{2}-(t^{2}-t^{2}_{0})[dr^{2}+\sinh^{2}r(d\theta^{2}+
\sin^{2}\theta d\varphi^{2})]-\frac{t^{2}}{t^{2}-t^{2}_{0}}d(x^{5})^{2},\ \
(|t|\geq t_{0}),
\end{equation}
has been obtained in \cite{r10,iv} too.
Note that all these  solutions will be valid for the case of a
timelike 5-th coordinate
(after changing the sign).

\section{Conformal transformations and equations of state}

We restrict our attention to conformal factors of the form
\begin{equation}\label{conf}
{\cal F}(\varphi)=\varphi^{2n},
\end{equation}
Take the metrics of the form
\begin{equation}\label{cftr}
g_{\mu\nu}=\tilde{g}_{\mu\nu}\varphi^{-2n}.
\end{equation}
as observable ones.
We consider models with flat space sections separately. The metric (\ref{flat1})
after the transformation (\ref{cftr}) takes the following form:
\begin{equation}\label{4flat1}
\left\{
\begin{array}{ll}
ds^{2}_{n}=d\tau^{2}-
\tau^{2n/n-1}[dr^{2}+r^{2}(d\theta^{2}+\sin^{2}\theta d\varphi^{2})]&
,\ \mbox{if}\  n\neq1;\\
ds^{2}_{1}=d\tau^{2}-
e^{\pm2\tau}[dr^{2}+r^{2}(d\theta^{2}+\sin^{2}\theta d\varphi^{2})]&
,\ \mbox{if}\  n=1,
\end{array}
\right.
\end{equation}
where $\tau$ is the world time. Similarly one can derive an observable
metric for the solution (\ref{flat2})
\begin{equation}\label{4flat2}
\left\{
\begin{array}{ll}
ds^{2}_{n}=d\tau^{2}-
\tau^{2(n+1)/(n+2)}[dr^{2}+r^{2}(d\theta^{2}+\sin^{2}\theta d\varphi^{2})]&
\mbox{if}\  n\neq-2;\\
ds^{2}_{-2}=d\tau^{2}-
e^{\pm\tau}[dr^{2}+r^{2}(d\theta^{2}+\sin^{2}\theta d\varphi^{2})]&
\mbox{if}\  n=-2.
\end{array}
\right.
\end{equation}

Let us use the well known expression for the matter energy-momentum tensor in GR
\begin{equation}\label{mat}
T_{\mu\nu}=(p+\varepsilon)u_{\mu}u_{\nu}-pg_{\mu\nu}
\end{equation}
In our case there is no nongeometric matter, however there is the scalar field
$\varphi$, and  we shall consider it to be describing external matter.
Substituting the metric components from (\ref{4flat1})--(\ref{4flat2}) to
the left-hand side of the Einstein equations and $T_{\mu\nu}$ from (\ref{mat})
to right-hand side, we can deduce expressions for the energy density $\varepsilon$
and the pressure $p$ of the effective simulated matter in comoving reference frame:
\begin{equation}
\varepsilon=3\dot{\lambda}^{2};\ \ \ p=-3\dot{\lambda}^{2}-2\ddot{\lambda}^{2},
\end{equation}
where $\dot{\lambda}\equiv d\lambda/d\tau$. Let us introduce $k$ which
determines the equation of state:
\begin{equation}\label{steq}
p=k\varepsilon.
\end{equation}
For the flat space section $k$ is constant. For the metrics (\ref{4flat1}) we
have
\begin{equation}
k=-1+\frac{2}{3}\cdot\frac{n-1}{n},
\end{equation}
for instance,
\[
\begin{array}{llll}
\mbox{if}\ n=1&\rightarrow& p=-\varepsilon&-\mbox{Zeldovich matter;}\\
\mbox{if}\ n=-2&\rightarrow& p=0&-\mbox{dust;}\\
\mbox{if}\ n=-1&\rightarrow& p=\varepsilon/3&-\mbox{radiation;}\\
\mbox{if}\ n=-1/2&\rightarrow& p=\varepsilon&-\mbox{stiff matter;}
\end{array}
\]
For the metrics (\ref{4flat2}) in a similar manner we find
\begin{equation}
k=-1+\frac{2}{3}\cdot\frac{n+2}{n+1},
\end{equation}
For instance,
\[
\begin{array}{llll}
\mbox{if}\ n=-2&\rightarrow& p=-\varepsilon&-\mbox{Zeldovich matter;}\\
\mbox{if}\ n=1&\rightarrow& p=0&-\mbox{dust;}\\
\mbox{if}\ n=0&\rightarrow& p=\varepsilon/3&-\mbox{radiation;}\\
\mbox{if}\ n=-1/2&\rightarrow& p=\varepsilon&-\mbox{stiff matter;}
\end{array}
\]

So, for any equation of state (any $k$=const) in both solutions
one can select an appropriate value of the index $n$ for the conformal factor and vice versa,
i.e. the 5-d solutions (\ref{flat1}) and (\ref{flat2}) describe all cosmological
4-d models of Friedmann type with a flat 3-d space section and the matter
equation of state
$p=k\varepsilon$. This result
is in accordance with those, obtained in \cite{r7} by another method.

\section{Nonflat cosmological models}

For the solutions (\ref{close}) and (\ref{open}), corresponding to cosmological
models with nonflat 3-d space sections, the 4-d metrics after the conformal
transformation can be written in the form
\begin{equation}
A)\ \ ds^{2}_{n}=\frac{(t_{0}^{2}-t^{2})^{n}}{t^{2n}}dt^{2}-
\frac{(t^{2}_{0}-t^{2})^{n+1}}{t^{2n}}[dr^{2}+\sin^{2}r(d\theta^{2}+
\sin^{2}\theta d\varphi^{2})],
\end{equation}
where the proper time $\tau$ can be found from the expression
\begin{equation}\label{ftime1}
\tau=\pm\int\frac{(t_{0}^{2}-t^{2})^{n/2}}{t^{n}}dt+\tau_{0};
\end{equation}
\begin{equation}
B)\ \ ds^{2}_{n}=\frac{(t^{2}-t^{2}_{0})^{n}}{t^{2n}}dt^{2}-
\frac{(t^{2}-t^{2}_{0})^{n+1}}{t^{2n}}[dr^{2}+\sinh^{2}r(d\theta^{2}+
\sin^{2}\theta d\varphi^{2})],
\end{equation}
where
\begin{equation}\label{ftime2}
\tau=\pm\int\frac{(t^{2}-t^{2}_{0})^{n/2}}{t^{n}}dt+\tau_{0};
\end{equation}

Simple reasoning provides the following expression for the effective
$\varepsilon$ and $p$
\begin{equation}
\varepsilon=3\dot\lambda^{2}+3se^{-2\lambda};\ \
p=-3\dot\lambda^{2}-2\ddot\lambda-se^{-2\lambda}.
\end{equation}
In both these cases the coefficient $k$ is
\begin{equation}
k=\frac{(2n+1)t^{2}-(n+2)nt^{2}_{0}}{3[(2n+1)t^{2}+n^{2}t_{0}^{2}]},
\end{equation}
i.e. is in general a function of time. There are some
special cases, when $k=const$:

а)  $n=0$ or $-1$ $\rightarrow k=1/3$ --- radiation;

b)   $n=-1/2 \  \  \rightarrow k=1$ --- stiff matter.
In all other cases $k$ is a function of time. Below we comment on
some peculiarities
of 8 variants, depending on values of $n$:

a) each variant contains a time interval, when the Universe is expanding;

b) for most of the variants matter passes through a dust-like state;

c) for $n\in(0,1)$ and $n\in(-2,-1)$ in closed models there is an infinite
expansion for a finite time;

d) there is no variants that could give the sequence matter states:
vacuum --- radiation --- dust. For instance, open models for $n\ge1$
contain the sequence: vacuum --- dust --- radiation and for $n\in(0,1)$
closed models give the sequence: radiation --- dust --- vacuum;

e) when $n\le-1/2$, for open models the function $k(t)$ acquires vertical
asymptotes and has the corresponding peculiar moments of time. At these moments
a specific kind of "phase transition" takes place;

f) cases with $n\le-1$ contain examples of cosmological models without
cosmological singularities.


\section{General conformal transformation}

Results that have been obtained in the previous section, demonstrate pros and
cons of the interpretation, proposed here. As since Big Bang matter
has evolved and passed through a number of a different states it would be
reasonable to create  4-d cosmological solutions with a variable equation of state
that would have in this case a phenomenological nature.
The application of a conformal transformation, on the other hand, gives a variable
equation of state automatically, but there appears the problem of choosing the conformal factor.
Conformally transformed 5-d solutions with nonzero space curvature,
as have been shown in previous section, give neither the required sequence
of matter states, nor  equations of state of the form $p=k\varepsilon$
with arbitrary $k=const$.

To obtain this class of equations of state, it is necessary to
abandon the  restriction (\ref{conf}) and consider the case of a general
conformal transformation $\cal F(\varphi)$:
\begin{equation}
\tilde g_{\mu\nu}=\frac{1}{{\cal F}^{2}(\varphi^{2})}g_{\mu\nu}.
\end{equation}
Then for close models we have
\begin{equation}
ds^{2}={\cal F}^{2}(\varphi^{2})dt^{2}-{\cal F}^{2}(\varphi^{2})(t_{0}^{2}-
t^{2})dl^{2}_{+};\ \ \varphi^{2}=\frac{t^{2}}{t_{0}^{2}-t^{2}}.
\end{equation}
In the same way for open models
\begin{equation}
ds^{2}={\cal F}^{2}(\varphi^{2})dt^{2}-{\cal F}^{2}(\varphi^{2})(t^{2}-
t^{2}_{0})dl^{2}_{-};\ \ \varphi^{2}=\frac{t^{2}}{t^{2}-t^{2}_{0}}.
\end{equation}
Here $l_{+}$ and $l_{-}$ are the 3-d intervals of closed and open models.
A calculation, performed in the same way as in section 3, leads to the
following expression for $k$:
\begin{equation}
k_{+}=\frac{p}{\varepsilon}=
\frac{4t_{0}^{4}t^{2}({\cal F}'^{2}-2{\cal F}{\cal F}'')-
4{\cal F}{\cal F}'t_{0}^{2}(t_{0}^{4}-t^{4})+
t_{0}^{2}{\cal F}^{2}(t_{0}^{2}-t^{2})^{2}}
{12{\cal F}'^{2}t^{2}t_{0}^{4}-12{\cal F}{\cal F}'t^{2}t_{0}^{2}(t_{0}^{2}-
t^{2})+3{\cal F}^{2}t_{0}^{2}(t_{0}^{2}-t^{2})^{2}}.
\end{equation}
$k_{-}$ for open model can be derived from $k_{+}$ by changing
$t\rightarrow it$, $t_{0}\rightarrow it_{0}$, $\varphi^{2}\rightarrow-\varphi^{2}$:
\begin{equation}
k_{-}=
\frac{4t_{0}^{4}t^{2}({\cal F}'^{2}-2{\cal F}{\cal F}'')+
4{\cal F}{\cal F}'t_{0}^{2}(t_{0}^{4}-t^{4})+
t_{0}^{2}{\cal F}^{2}(t_{0}^{2}-t^{2})^{2}}
{12{\cal F}'^{2}t^{2}t_{0}^{4}+12{\cal F}{\cal F}'t^{2}t_{0}^{2}(t_{0}^{2}-
t^{2})+3{\cal F}^{2}t_{0}^{2}(t_{0}^{2}-t^{2})^{2}}.
\end{equation}
Here ${\cal F}'=d{\cal F}(\varphi^{2})/d(\varphi^{2})$. Making a change of
the variable and the function,
$$
x=t^{2}/(t^{2}-t_{0}^{2}),\ \ u={\cal F}^{\frac{2-\sigma}{2}},
\ \ (\sigma=1-3k\neq2)$$
we obtain for open models the hyphergeometric Gauss
equation
\begin{equation}\label{hypherg}
x(x-1)u''+[(\alpha+\beta+1)x-\gamma]u'+\alpha\beta u=0,
\end{equation}
where $\alpha+\beta+1=(\sigma+1)/2;\ \ \gamma=1/2;\ \ \alpha\beta=
\sigma(\sigma-2)/16$. For closed models $x$ should be changed to $-x$.
The general solution of (\ref{hypherg}) has the form
\begin{equation}
u=C_{1}F(\alpha,\ \beta,\ 1/2,\ x)+C_{2}\sqrt{x}F(\alpha+1/2,\
\beta+1/2,\ 3/2,\ x)
\end{equation}
where $F$ is the  hyphergeometric function.

So, if geometrized matter has the equation of state  $p=k\varepsilon$,
where $k=const$, then the conformal factor $\cal{F}$ should have a
form more general then
(\ref{conf}) and should satisfy (\ref{hypherg}). For example,
for dust $(k=0)$ Eq. (\ref{hypherg}) has the particular solution
${\cal F}=1+\sqrt{1\pm\varphi^{2}}$ for closed and open models, respectively.
For Zeldovich matter $(k=-1)$ ${\cal F}=1\pm\varphi^{2}$. For $\sigma=2$
$(k=-1/3)$ we obtain a linear equation of second order for $\cal{F}$ and
it's solution can be written in terms of elementary functions.

\section {Conclusion}

So, in 5-d Kaluza-Klein theory, matter can be indeed described by a
geometric scalar field $(\varphi=\sqrt{G_{55}})$, while
in 4-d GR it is conventional to substitute an arbitrary through tensor $T_{\mu\nu}$.
The freedom in choosing the conformal factor corresponds to different possible equations
of state for matter. In the present paper it has been demonstrated
that for flat cosmological models the equations of state
is prescribed by the index of the conformal factor $\varphi^{n}$. For
nonflat models a number of variants is possible. Variable equations of
state with "phase transitions" of matter are obtained automatically.
To describe of matter with constant equations of state  it is necessary to
use conformal factors of a more general form than the exponential function.

\end{document}